\documentstyle[11pt,newpasp,twoside,epsf]{article}
\def\edcomment#1{\iffalse\marginpar{\raggedright\sl#1\/}\else\relax\fi}
\reversemarginpar

\begin{document}
\title{Keck adaptive optics observations of TW Hydrae Association members}
\author{Bruce Macintosh, Claire Max}
\affil{Lawrence Livermore National Lab, Livermore, CA 94551}
\author{Ben Zuckerman, Eric E. Becklin, Denise Kaisler, Patrick Lowrance, Alycia Weinberger}
\affil{UCLA, 405 Hilgard Ave., Los Angeles, CA 90095}
\author{Julian Christou}
\affil{Center for Adaptive Optics, UCSC, Santa Cruz, CA 95064}
\author{Glenn Schneider}
\affil{Steward Observatory, U.Arizona, Tucson, AZ 85721}
\author{Scott Acton}
\affil{California Association for Research in Astronomy, Kamuela, HI 96743}

\begin{abstract}

Adaptive optics (AO) on 8-10 m telescopes is an enormously powerful tool
for studying young nearby stars. It is especially useful for searching for
companions. Using AO on the 10-m W.M. Keck II telescope we have measured
the position of the brown dwarf companion to TWA5 and resolved the primary 
into an 0.055\arcsec\ double. Over the next several years follow-up astrometry
should permit
an accurate determination of the masses of these young stars. We have also
re-observed the candidate extrasolar planet TWA6B, but measurements of its
motion relative to TWA6A are inconclusive. We are carrying out a search
for new planetary or brown dwarf companions to TWA stars and, if current
giant planet models are correct, are
currently capable of
detecting a 1 Jupiter-mass companion at $\sim1.0\arcsec$ and a 5 Jupiter-mass
companion at $\sim0.5\arcsec$ around a typical TWA member.

\end{abstract}

\section{Adaptive optics capabilities and limitations}

Adaptive optics is a very powerful tool, allowing large ground-based
telescopes to reach their diffraction-limited resolutions in the near IR -- 
resolutions of $0.03\arcsec$ or better. For some applications, this allows
ground-based telescopes to perform as well as or better than 
the 2.5 m HST. Detection
of faint companions is one such application, since by concentrating the
light of a companion into a diffraction-limited point, an AO system
can enhance the contrast of such a companion
by a factor of 100 or more relative to non-AO observations. 

Young nearby stars such as those in the TW Hydrae association are an ideal
target set for an AO companion search. They are still young enough that
planetary-mass companions are detectable (Burrows et al. 1997) and close
enough, in comparison to regions such as Taurus, that these companions can
be seen at separations of 20-50 AU. TWA members are also
generally bright enough at visible wavelengths to give good AO performance
even for high-order AO systems such as Keck.

\section{Keck adaptive optics}
The Keck AO system is described in Wizinowich et al. (2000a and 2000b). 
The observations discussed here were made with KCam, 
an interim near-infrared camera provided by UCLA. KCam has a 
$256 \times 256$ pixel NICMOS3 HgCdTe array. 
Warm magnifying optics provide a 
plate scale of 0.01747 \arcsec/pixel. A cold filter
wheel selects the standard J, H, and K' bands; an external warm filter wheel
holds additional narrowband and neutral density (ND) filters. The latter are 
required to avoid saturation on any star brighter than $m_{H}\approx8$.

\section{Observations}
We observed several TWA members in February 2000, both to confirm candidate
companions from previous NICMOS observations and to search for new
substellar companions. Results will be presented in detail in future
papers. Here we discuss our observations of TWA5, TWA6, and our sensitivity
estimates for new companion detection.

\subsection{TWA5}
TWA5 was observed to measure the current separation of the brown dwarf
TWA5B (Lowrance et al. 1999, Webb et al. 1999). Seven images were taken, each of 10 coadds of
0.62 seconds integration. Two (in which TWA5A saturated) 
were with a standard H-band filter, five with
the H filter plus an ND1 (90\% attenuation) filter. A typical image is shown
in Figure 1a.

\subsubsection{TWA5A}
The most noteworthy discovery made was that TWA5A is a $\sim 0.06\arcsec$
\ binary
(Figure 1b), easily visible in the unsaturated ND-filter images. TWA5A was 
thought to be a spectroscopic binary (Webb et al. 1999), 
although no period solution
has yet been found (Torres et al. 2001.) 
Since our detection implies a period of several years, it is likely 
that the system is further multiple (making, including the brown dwarf, at
least a quadruple) and one of the two components we detect is itself a 
short-period binary.
\begin{figure}
\plottwo{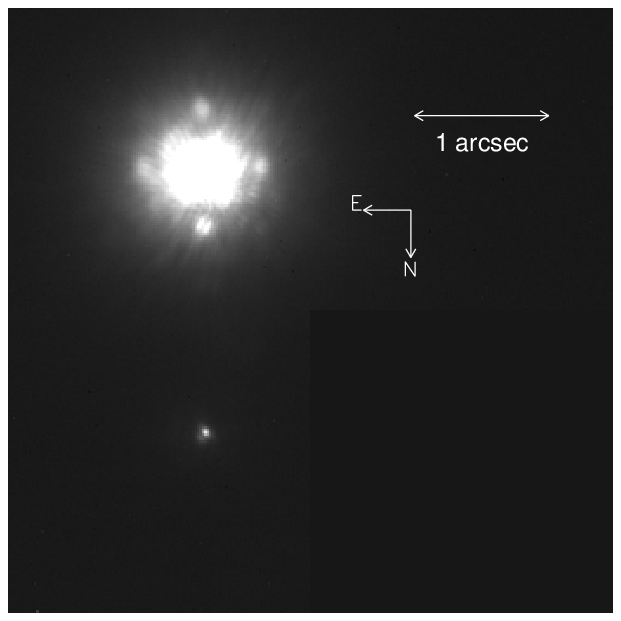}{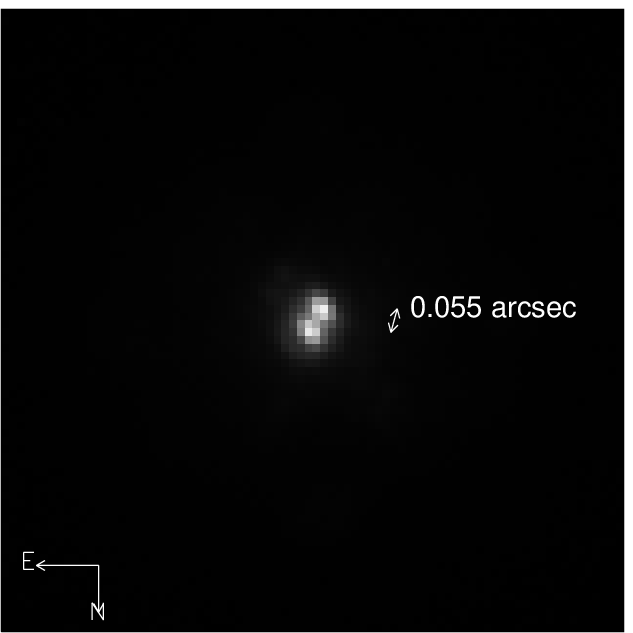}
\caption{\textbf{Left:}Keck AO images of TWA5A (top, highly saturated) and B
\textbf{Right:} Close-up of TWA5Aa and Ab. Both are 6.25 second H-band images. Right-hand image taken through a ND1 filter.}
\end{figure}

The pair is close enough to require careful deconvolution to obtain
astrometry and
photometry. Fortunately the brown dwarf TWA5B is present in all the
images (even those taken through the ND filter) and this provides an
estimate of the PSF.  The data were reduced with the multi-frame blind
deconvolution algorithm, 
IDAC\footnote{http://babcock.ucsd.edu/cfao\_ucsd/idac/idac\_package/idac\_index.html}
(Jefferies \& Christou, 1993) using
the 5 unsaturated images simultaneously and PSF estimates as discussed
above.  The procedure followed that described in Barnaby et al.
(2000) using deconvolution of upsampled data and and gaussian fitting of the 
results to obtain
the photometry and astrometry of the binary. 
We derive $\Delta H=0.09\pm0.02$ with the brighter component (TWA5Aa) being the
southernmost, a separation of $0.0548\pm0.0005$ arcseconds, and a position
angle from Aa to Ab of $25.9\pm0.5$ degrees.

TWA5A was unresolved in NICMOS acquisition images obtained on April 25 1998. 
Although the resolution of NICMOS is less than the Feb. 2000 separation of
the pair, this separation would still have been detectable as an elongation of
the PSF or in template subtractions. 
Analysis of the NICMOS images indicates at the 3-sigma confidence level 
that the separation at this epoch must have been less than 0.025\arcsec,
further evidence of a short orbital period. Given the presence of 
TWA5B to serve as an absolute astrometric reference, and the possibility of
obtaining resolved spectra with adaptive optics systems, 
an unambiguous mass solution should be obtainable 
for TWA5A over the next several
years. This in turn can be compared to PMS evolution models to validate
evoluationary tracks. Such comparison will also better constrain
the age of the system, and in turn the new age can be used for a mass
determination of TWA5B.

\subsubsection{TWA5B}
The brown dwarf TWA5B was easily detected, even in images
taken with the neutral density filter.
We measured a separation between it and the photocenter of the TWA5A pair
of $1.956\pm0.012$ arcseconds, with the uncertainties being dominated
by the distortions mentioned below. This indicates very little change since
the measured NICMOS value of $1.9639\pm0.011$ arcseconds in Lowrance et al. 
1999. TWA5B is further discussed in Lowrance et al. (2001) in this volume.

\subsection{TWA6}
The most interesting candidate companion discovered during the NICMOS
Environments of Nearby Stars (EONS) survey was the object which we will
tentatively designate TWA6B. This is a $m_{H}=20$
companion 2.5\arcsec \  from the TWA member TWA6, discussed in Lowrance
et al. 2001 in this volume.

On Feb 21 2000 we
took a series of images of TWA6 with Keck AO. Short-exposure images were 
taken with the ND filter to acquire and align TWA6 itself and then
without the ND filter to search for the companion. TWA6B was detected in
each of four 120 second H-band exposures.

%\begin{figure}
%\plotone{fig3.eps}
%\caption{Keck 120 second H-band image of TWA6AB. 
%Image has been high-pass filtered to
%surpress the scattered light halo of TWA6A. Candidate planetary companion
%TWA6B is marked with a
%circle. MAYBE OMIT THIS FIGURE TO SAVE SPACE????}
%\end{figure}

We used these to measure the change in separation between TWA6B and TWA6A.
The NICMOS measurement on 20 May 1998 was $2.549\pm0.011$ arcseconds. 
TWA6 has a proper motion from Tycho 2 of $57.0\pm3.0$ mas/year west and
$20.6\pm2.8$ mas/year south, almost directly towards TWA6B, 
so the primary effect of its 
motion if TWA6B is a background object should be to decrease the separation
to $2.459\pm0.012$ arcseconds . Since
the orientation of our Keck images is influenced by uncertainty in the
position of the image derotator used, we will only discuss measurements of the
separation here.  

Unfortunately, TWA6A is of course
saturated in the long exposure non-ND images while TWA6B is undetectable
in the unsaturated images.  The neutral density filters, 
however, induce a position shift in the image of a star; this position
shift was measured once later in the night using an intermediate-brightness
star, but it is unknown if this shift varies due to small changes in the
position of the filter as it is moved in and out. 
Including the estimated shift, 
we measure a separation between TWA6 and TWA6B of $144.1\pm0.5$ pixels.

The plate scale of the camera was determined through measurements of the 
Hipparcos binaries HIP51802, HIP50223, HIP52913
and HIP48618. The derived plate scale is 
$0.01747\pm0.00006$ arcseconds per pixel. Astrometric calibration will be
discussed in detail in a seperate paper. Unfortunately, the data available
are insufficient to measure any distortions inherent in the camera's optics or
due to the ND filter. The TWA5B measurement discussed above is in good 
agreement
with previous measurements, and the scatter between different HIP stars is
consistent with the uncertainties in their positions, 
indicating that distortions are small, but 
the ND filter effects are unknown. 

This leads to a separation in February 2000 of $2.517\pm0.012$ arcseconds. 
This is two sigma away from the measured NICMOS position and approximately
3 sigma away from the position of a background star - still inconclusive.
(Figure 2.) Two additional measurements made in April 2000, with somewhat
greater uncertainties, are closer to the background star position but also
inconclusive. The February and April measurements disagree enough to indicate
that an unmodeled source of error - almost certainly uncertainties in the
displacement due to the ND filter - is present. Measurements made in February
2001, with a better camera and a longer time baseline, 
should determine the true nature of TWA6. 

\begin{figure}
\plotone{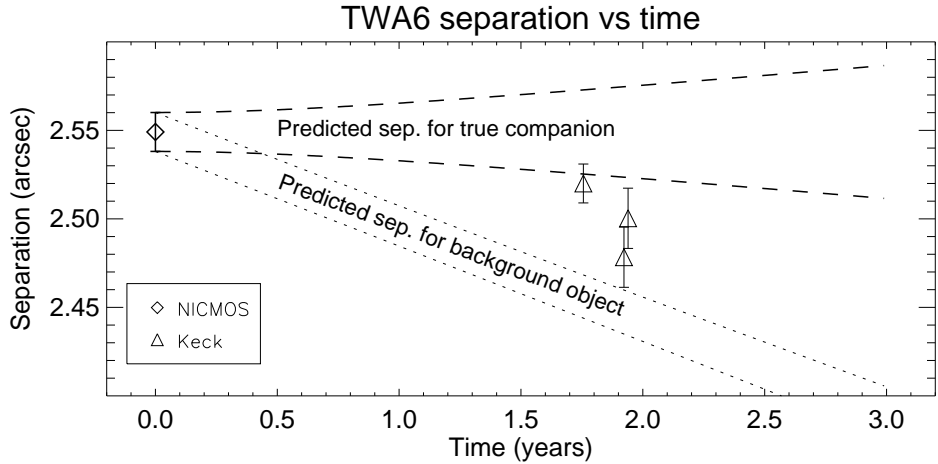}
\caption{ Separation vs time for TWA6AB, showing the original 1998 NICMOS
measurement and the 2/2000 and 4/2000 Keck measurements. Diagonal lines
show the likely range for a background object, and the slightly curved
horizontal lines show the 1-sigma error on the initial position measurement
plus the allowable orbital motion for a bound companion.}
\end{figure}

\section{Search for new companions}
Even if TWA6 proves to be a background star, it does demonstrate that we have
the necessary sensitivity to see planet-like companions to nearby young stars
at interesting $(\sim 100 AU)$ separations. We have begun a program to survey
the ever-increasing number of known members of TWA and other young groups,
as well as other young field
stars accessible to the Keck observatory. 
Analysis of these data are ongoing. Figure 3 shows an estimate of our typical
sensitivity, compared to that of earlier NICMOS measurements. 
The sensitivity in both cases has been computed based on the variance of
PSF-subtracted data in successive annuli. This should not be considered 
definitive - a true detection would require multiple images of a given target
at the 5-sigma level - but the variances were calculated in the same way for both instruments,
and illustrate that the companion-detection sensitivity of Keck AO is currently
comparable to NICMOS. Keck AO sensitivity at large radii was limited by 
high readnoise in the KCam camera and has improved as new instruments are
used.

\begin{figure}
\plotone{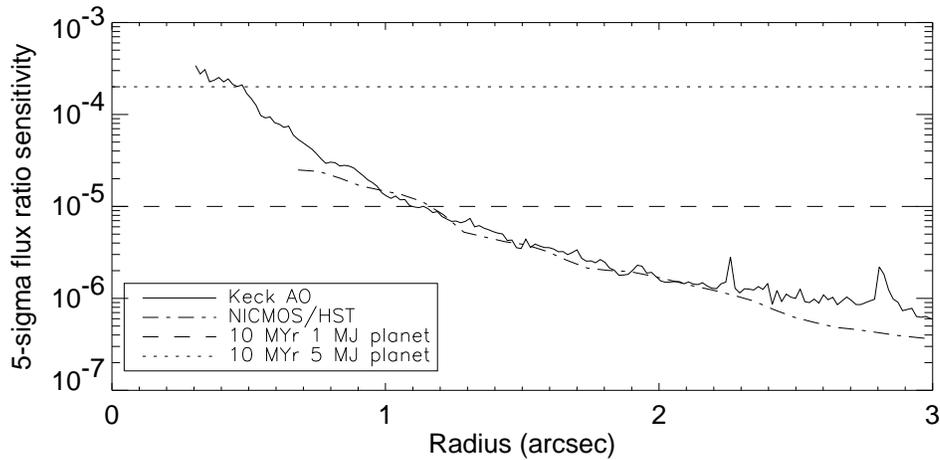}
\caption{Estimated sensitivity to faint companions in Keck AO imaging.
For comparison, a similarly-calculated sensitivity for NICMOS is shown. 
Overlaid are the flux-ratios for 1 and 5 Jupiter-mass companions to TWA stars,
from Burrows et al. 1997}
\end{figure}

Improved image processing techniques
are expected to enhance sensitivity by a factor of 3-10 at small radii.
Even now, we are capable of detecting a $1 M_{J}$ companion at a separation of
$1\arcsec$ and a $5 M_{J}$ companion at a separation of $0.5\arcsec$,
assuming a TWA age of 10 MYr and using the models of Burrows et al. (1997).
As more AO systems
become operational on 8-10m telescopes, particularly in the southern hemisphere,
direct detection of an extrasolar planet orbiting a young star seems almost
inevitable, if such massive planets do exist with orbital separations
greater than 20 AU.

\acknowledgements
We would like to thank the Keck AO team, Peter Wizinowich, Olivier Lai,
and Paul Stomski, for assistance with observations. 
This research was performed under the auspices of the U.S.
Department of Energy by Lawrence Livermore National Laboratory 
under Contract W-7405-ENG-48, and
also supported in part by the Center for Adaptive Optics under
the STC Program
of the National Science Foundation, Agreement No. AST-9876783

\end{document}